\documentstyle[aps,preprint]{revtex}

\begin{document}
\draft
\title{ Finite Temperature Fractional Quantum Hall Effect }
\author{
Lizeng Zhang }
\address{
Department of Physics and Astronomy \\
The University of Tennessee, Knoxville, TN 37996\\
and Solid State Division, Oak Ridge National Laboratory\\
Oak Ridge, Tennessee 37831\\
}
\maketitle
\begin{abstract}
We investigate fractional quantum Hall effect at finite
temperature using a fermion Chern-Simons field theoretical
approach.  In the absence of impurity scattering,
the essential aspects of fractional quantum Hall effect,
such as the quantization of
Hall conductance, as well as quasi-particle charge and statistics
are unrenormalized by thermal fluctuations. On the other hand, we find
that the low energy excitation spectrum at finite $T$ may undergo some
qualitative changes as temperature raises. Possible experimental
consequences are discussed.
\end{abstract}
\pacs{PACS numbers: 73.40.Hm }

{\sl Introduction} --
The novel phenomenon of fractional quantum Hall effect (FQHE)
may be understood theoretically as a manifestation of certain
two dimensional (2D) highly correlated fermion states
\cite{Laughlin,book}. Most of the theoretical efforts thus far
have been given to the study of
zero temperature properties, where the ground state is known to be
incompressible, separated from higher energy states by a
gap $\Delta$ produced by strong electron-electron interaction.
On the other hand, experimental measurements are done, of course,
at finite temperature.  But from the above mentioned property
of the ground state, which has been established
firmly through more than a decade of extensive study, it is
reasonable to expect that at finite
temperature the effect of thermal fluctuations would not be important,
as long as $kT \leq \Delta$.

However, for a quantitative comparison between theory and experiment,
there are several issues that need to be addressed concerning
the finite (low) temperature properties of the FQHE:
the first one is the fundamental question about the accuracy of the FQHE
at non-zero $T$. How do thermal fluctuations affect the
quantization of the Hall conductance? Another
important question is
the $T$ dependence of the quasi-particle gap which can be explicitly
computed at $T=0$ in Laughlin's theory
\cite{Laughlin,book} for the `fundamental states' and in Jain's
composite fermion theory \cite{Jain} for general filling fractions,
and can be implicitly measured experimentally through the $T$
dependence of the longitudinal resistivity \cite{book,Du}.
In this paper we study FQHE at finite temperature.
In particular, we investigate the $T$ dependence of low energy
collective excitations. At zero $T$, the existence of a
special roton-like excitation in the Laughlin states was first
demonstrated by Girvin et al
through the Feynman-Bijl approach \cite{GMP}, and later by
Zhang et al using a Chern-Simons Landau Ginzburg theory \cite{CSLG}.
For general filling fraction $\nu$, this problem was investigated
within the composite fermion picture aided with the
method of `Chern-Simons transformation' \cite{LF}-\cite{HSH}.
Our approach is a fermion Chern-Simons field theory formalism, which
has been used previously by Lopez and Fradkin \cite{LF} to study FQHE
at zero $T$, and by Halperin, Lee and Read in their study of the
$\nu=1/2$ state \cite{HLR}. Presented in an equivalent quantum
many-body language, Simon and co-workers \cite{SH,HSH} have studied
this Chern-Simons--composite fermion approach in great detail.
An important point addressed in Ref.\cite{HLR}-\cite{HSH} is
the problem of mass renormalization of composite fermions.
Without treating it properly, this fermion Chern-Simons theory
approach will set the system at a spurious energy scale
$\hbar\tilde{\omega}_{c}$
(see below), which is on the order of (although less than) the
bare cyclotron energy, while the true physical energy scale
of the problem is given by electron-electron interactions.
However, the question which concerns us here is the
{\em temperature dependence} of the
collective excitations rather than evaluation of the precise value
of the energy gap. Thus while the issue of mass renormalization
is important and its phenomenological Fermi-liquid treatment
(discussed in Refs.\cite{HLR}-\cite{HSH}) may be straight
forwardly generalized to the finite $T$ case, it will not be
consider here. This issue, along some details of the present
work, will be discussed somewhere else \cite{LZ}.

{\sl Approach} --
In the spirit of composite fermion approach, we describe 2D
electrons of band mass $m_b$ in a magnetic field at temperature
$kT=1/\beta$ by a coherent functional integral with action
(in the unit $\hbar=1$):
\begin{eqnarray}\label{action}
S=\int_{0}^{\beta}d\tau\int d^{2}r \{
\psi^{\dagger}({\partial_\tau}-ia_{0}-\mu)\psi+
{1\over{2m_b}}|(-i{\vec \nabla}+\frac{e}{c}{\vec A}-{\vec a})\psi|^{2}
+{i\over {4\pi\tilde{\phi}}}\epsilon_{\mu\nu\lambda}
a_{\mu}{\partial}_{\nu}a_{\lambda}  \} \nonumber \\
+{1\over 2}\int_{0}^{\beta}d\tau \int d^{2}r\int d^{2}r^{\prime}
\psi^{\dagger}\psi(r) v(\vec{r}-\vec{r}\;^{\prime})
\psi^{\dagger}\psi(r^{\prime}) \;\;\;,
\end{eqnarray}
where we attached even flux quanta per fermion through the well-known
`Chern-Simons term' with statistical gauge field $a_\mu$.
In the above expression, ${\tilde \phi} = 2p$, and $p$
is an integer.
${\vec A}$  is the vector potential for the magnetic field
and the chemical potential $\mu$ fixes the Landau level (LL)
filling fraction at $\nu$. $v(\vec{r})$ is a two-body interaction
potential. For Coulomb interaction, $v(\vec{r})=e^{2}/\epsilon r$,
where $\epsilon$ is the dielectric constant.
A similar Euclidean action has been studied in Ref.\cite{RSS} in the
context of anyon superconductivity.

While this action is equivalent to the usual one without the
statistical gauge field, it provides us a convenient starting point
for approximations: consider the homogeneous {\it liquid}
saddle point solution for the Chern-Simons field
$a_{\mu}=\bar{a}_{\mu}$, such that
\begin{equation}\label{mfcs}
|\vec{\nabla} \times \vec{\bar{a}}| = 2\pi\tilde{\phi}\bar{\rho}
\;\;, \;\;\;\;\;\; \bar{a}_{0} = 0 \;\;,
\end{equation}
where $\bar{\rho}$ is the average particle density. In this mean
field theory, an electron feels an effective magnetic field
\begin{equation}\label{Beff}
B_{eff} \equiv |\vec{\nabla} \times \vec{A}_{eff}| \equiv
|\vec{\nabla} \times (\vec{A} - \frac{c}{e}\vec{\bar{a}})|
\;\;\;.
\end{equation}
For the filling fraction
$\nu$ such that $1/(1/\nu-2p) = n$, an integer,
the mean-field theory possesses a ground state of $n$ filled LL,
which is thus incompressible and stable against weak perturbations.
The filling fraction $\nu$ which
satisfies this condition is precisely the value where FQHE
arises. This correspondence between the fractional and
the integer QHE is precisely the basic idea of Jain's
composite fermion theory \cite{Jain} which underlines the
present approach. Fluctuation corrections may be systematically
build around this (stable) saddle point solution \cite{LF}.

At finite $T$, we shall also take such a homogeneous saddle point
solution as our starting place and consider only fluctuation
corrections on the one-loop level.
The implicit assumption is that at thermal equilibrium the
composite fermion picture remains a good description,
hence effects of the {\it thermal disintegration} of
composite fermions, i.e., the unbinding of particles and
(even number of) vortices are not important for the
temperature range under consideration.
Experimentally it has been found \cite{Willett} that the $\nu=1/2$
anomaly persists to some temperature range in which
the FQHE at other filling fractions are smeared
out by thermal fluctuations, which suggests strongly \cite{Willett} that
composite fermions \cite{HLR} are rather robust against thermal
fluctuations, and the disintegration temperature
$T_d$ of composite fermions lies well above the (zero $T$)
energy gap of the incommpressible FQH states.

With Eqns.(\ref{mfcs}) and (\ref{Beff}), the mean field action
is given by
\begin{equation}
S_{0} = \int_{0}^{\beta}d\tau\int d^{2}r \left \{
\psi^{\dagger}({\partial_\tau}-\mu)\psi+
{1\over{2m_b}}|(-i{\vec \nabla}+\frac{e}{c}{\vec A_{eff}})\psi|^{2}
\right \} \;\;\;.
\end{equation}
This action describes a system of non-interacting fermions with
magnetic field $B_{eff}$, in which the energy spectrum is given by the
effective LL $\epsilon_{l}=\tilde{\omega}_{c}(l+\frac{1}{2})$,
where $\tilde{\omega}_{c} = eB_{eff}/m_{b}c$, is the effective cyclotron
frequency.
The chemical potential $\mu$ is determined by the condition \cite{RSS}
\begin{equation}\label{chemp}
n=\sum_{l}f(\epsilon_{l}-\mu)\;\;, \;\;\;\;\;
f(\epsilon_{l}-\mu) = \frac{1}{e^{\beta(\epsilon_{l}-\mu)} + 1}
\;\;\;.
\end{equation}

To study the Gaussian fluctuations in the Chern-Simons gauge
field, we adopt the approach of \cite{HLR,SH} by taking a
transverse gauge such that $\vec{a}_{T} \parallel \hat{y}$,
$\vec{A}_{T} \parallel \hat{y}$ and choose $\vec{q} \parallel \hat{x}$.
In this case, $j_{x}(q,\omega)$ is simply given by
$\frac{\omega}{q}\rho(q,\omega)$.
Shifting $a_\mu$ by $\bar{a}_{\mu}$, i.e.,
$a_{\mu} \rightarrow \bar{a}_{\mu} + a_{\mu} $,
we can express the kernel of the Gaussian action for the statistical
field $a_\mu$ in terms of a $2\times 2$ matrix. At finite
frequency, it is given by
\begin{equation}
D^{-1}=\frac{1}{2\pi\tilde{\phi}}\left(
\begin{array}{cc}
\tilde{n}q^{2}/\tilde{\omega}_{c}\Sigma_{0}&iq(1-\tilde{n}\Sigma_{1})\\
-iq(1-\tilde{n}\Sigma_{1})&\tilde{\omega}_{c}
(\frac{\mu_s}{2p}\tilde{l}_{0}q+\tilde{n}(\Sigma_{2}+1))
\end{array}\right) \;\;\;,
\end{equation}
with $\tilde{n}=2pn$ and
\begin{eqnarray}\label{sigmaj}
\Sigma_{j}=\frac{e^{-x}}{n}\sum_{m<l}
\frac{l-m}{(\frac{i\omega_n}{\tilde{\omega}_{c}})^{2}-(l-m)^{2}}
\frac{m!}{l!}x^{l-m-1}
\{ f(\epsilon_{m}-\mu)-f(\epsilon_{l}-\mu) \} \nonumber \\
\left [ L_{m}^{l-m}(x) \right ]^{2-j}
\left [ (l-m-x)L_{m}^{l-m}(x)+2x\frac{dL_{m}^{l-m}(x)}{dx} \right ] ^{j}
\;\;\;.
\end{eqnarray}
In the above equation, $i\omega_{n}$ is the Matsubara frequency,
$L_{m}^{l}$ is the Laguerre polynomial, $x=(\tilde{l}_{0}q)^{2}/2$,
and $\tilde{l}_{0}=\sqrt{ c/eB_{eff}}$ is the effective magnetic length.
$\mu_{s}$ is the ratio of $\tilde{l}_{0}$ to the Bohr radius
$a_{0}=\epsilon /m_{b}e^{2}$. To obtain electromagnetic response,
we consider a fluctuation in $A_{\mu}$ such that
$A_{\mu} \rightarrow A_{\mu} + \delta A_{\mu}$.  Integrating out
the statistical field $a_\mu$, we arrive the effective action
\begin{equation}
{\tilde S}_{Gaussian}(\delta A_{\mu}) = \frac{1}{2}\sum_{q,i\omega_{n}}
\left(i\delta A_{0}(-q,-i\omega_{n})\;\;\;
\frac{e}{c} \delta A(-q,-i\omega_{n}) \right)
K(q,i\omega_{n})
\left ( \begin{array}{c} i\delta A_{0}(q,i\omega_{n}) \\
\frac{e}{c} \delta A (q,i\omega_{n}) \end{array} \right )
\;\;\;,
\end{equation}
with the kernel ($i\omega_{n} \neq 0$)
\begin{equation}
K(q,i\omega_{n})=\frac{n}{2\pi (det)}\left(
\begin{array}{cc}
\frac{q^{2}}{\tilde{\omega}_{c}}\Sigma_{0}&-iq\Sigma_{s}\\
iq\Sigma_{s}&\tilde{\omega}_{c}\Sigma_{r}
\end{array}\right)  \;\;\;.
\end{equation}
In the above equation, $\Sigma_{s} = \Sigma_{1}(1-\tilde{n}\Sigma_{1}) +
\tilde{n}\Sigma_{0}(\Sigma_{2}+1)$, $\Sigma_{r} = 1+\Sigma_{2}+
n\mu_{s}\tilde{l}_{0}q(\Sigma_{1}^{2}-\Sigma_{0}(\Sigma_{2}+1))$ and
$det=(1-\tilde{n}\Sigma_{1})^{2} -
\Sigma_{0}(n\mu_{s}\tilde{l}_{0}q+(\tilde{n})^{2}(\Sigma_{2}+1))$.
The electromagnetic response is obtained by the usual procedure
of substituting the Matsubara frequency $i\omega_{n}$
by $\omega - i\eta$. The above results differ from the zero $T$
calculations \cite{LF,SH} by the appearance
of the fermion distribution $f(\epsilon_{l}-\mu)$, hence the
necessity of summing up the additional LLs
in the expression for $\Sigma_j$.

{\sl FQHE at finite temperature} --
To answer the fundamental question concerning
the accuracy of FQHE at finite temperature, one has to consider
dissipative processes such as phonon or impurity scattering.
In the absence of them one should expect the same transport
properties at finite temperature as those at $T=0$.
In our calculation, this may be seen directly from the zero
$q$ response obtained from
\begin{equation}
\Sigma_{j}(q=0,\omega)=\frac{1}{n}\sum_{m=0}^{\infty}
\frac{m+1}{(\frac{\omega}{\tilde{\omega}_{c}})^{2}-1}
\{ f(\epsilon_{m}-\mu)-f(\epsilon_{m+1}-\mu) \} =
\frac{1}{(\frac{\omega}{\tilde{\omega}_{c}})^{2}-1} \;\;\;,
\end{equation}
where in the last equality we have used Eqn.(\ref{chemp}).
The zero frequency response needs to be calculated separately.
Straightforward calculation shows that at finite $T$ the
compressibility $\kappa$ acquires an exponential correction \cite{LZ}.
Now we consider the charge and statistics of the quasi-particle
(which is defined by excitations from the thermal equilibrium state
at a given $T$ \cite{PN}) at finite temperature.
Following Laughlin's gauge
argument \cite{book}, which can be readily generalized to the
finite $T$ case if one assumes states in each (pseudo) LL are
{\em uniformly} occupied with probability $f(\epsilon_{l}-\mu)$,
it is reasonable to expect the same fractionalization
of quasi-particle charge at finite temperature.
Indeed, if one lets $\delta A$ be a thread of unit flux
passing through the origin, one finds that total charge thus
induced is $\nu e$, independent of $T$. For $\nu = 1/(2p+1)$, we
equate this change to the charge of a quasi-particle, which
is in agreement with Laughlin's {\sl gedanken} experiment
\cite{Laughlin,book} at zero $T$.
In the general case, the charge of a quasi-particle may be obtained
by examining the the gauge invariant (finite temperature) one
particle Green function.
Following an argument similar to that of Fradkin \cite{Fradkin},
one obtains $eB_{eff} = e^{*}B$, which gives the quasi-particle
charge $e^{*} = e/(1+2pn)$. This is
the same as the zero temperature result \cite{Jain,Fradkin}.
Statistics of quasi-particles may be obtained from gradient
expansion of $\Pi^{0}$ \cite{Fradkin}. In the small $q$ limit, the
effective Chern-Simons coupling is given by
$1/\tilde{\phi}_{eff} = 1/\tilde{\phi}+n$. Counting the
original statistical phase of bare fermion, we find that
the statistical phase (relative to the boson) of quasi-particles
is $\pi(1-2p/(1+2pn))$.
All these results stated here, which involve only the physics
at large length scale, are $T$ independent and are given by those
obtained at $T=0$. This is a consequence of the fact
that at the long wavelength limit, temperature appears in
our formalism only through the sum $\sum_{l}f(\epsilon_{l}-\mu)$,
which is constant due to (\ref{chemp}). Also as a result of this,
the asymptotic behavior of quasi-particle amplitude distribution
remains the same as that at zero $T$, which can be readily
demonstrated by computing responses of the system to
point like external charge and external flux thread \cite{LF,SK,CQ}.
This result is significant, since the statistics of
quasi-particles described by the effective Chern-Simons
term obtained above corresponds to the phase gain from adiabatic
interchange of two particles separated {\em infinitely} apart.
Changes in the quasi-particle amplitude profile at finite $T$
will then lead to the corrections to the statistics of
quasi-particles \cite{SK} due to their overlap at finite distance.

{\sl Collective Excitations} --
Collective modes are obtained from the poles in $K(q,\omega)$
which describes the electromagnetic response of the system at
a given $T$.  At $q=0$, the cyclotron mode
$\omega_{c} = \frac{n}{\nu}\tilde{\omega}_{c}$
saturates the $f$-sum rule, in accordance with the Kohn's theorem
\cite{Kohn} which can be readily generalized to the finite $T$ case.
As pointed out in the introduction, the approach adopted here sets
a spurious energy scale $\hbar \tilde{\omega}_{c}$ in the problem.
Furthermore, once the saddle point solution Eqn.(\ref{mfcs})
is taken, the interaction $v(\vec{r})$ plays only a nominal role
in the perturbation expansion used here.  Since FQHE occurs
as a result of strong electron-electron correlations,
this artifact of our approach seems quite disturbing.
One can use the following rationale to justify the fermion
Chern-Simons field theory approach \cite{Jain}: although one
needs interaction to create a composite fermion (i.e., the bound
state of a bare fermion and $2p$ vortices), hence FQHE, once
it is formed the residual interaction among the composite
fermions becomes unimportant. Since the presence of
$v(\vec{r})$ does not change the qualitative physics and is
deceiving about the role of interaction in this
formalism, we shall set $\mu_s$ to zero hereafter.

Fig. 1 shows
the evolution of the lowest few branches of the collective modes
in the $\nu = 1/3$ case as $T$ is raised from zero.
At finite $T$, each branch of the zero temperature modes splits
into two, where the upper one of the two retains all the weight
at small $q$ value.
This splitting is due to the finite probability
of occupying higher (pseudo) LLs at non-zero $T$.
Inspecting the lowest mode, we see that the energy of the roton
minimum $\omega_{min}$ is rather insensitive to $T$, while
its position $q_{min}$ has stronger temperature
dependence.  This situation is depicted in Fig. 2.
In general, increasing $T$ causes red shift of $q_{min}$.
While its precise physical picture is unclear at the present time,
this red shift may be understood roughly as a result of weakening of
the roton-roton binding energy at small $q$ \cite{GMP} due
to thermal fluctuations.
A recent optical experiment measured directly the long wavelength
($q=0$) collective modes \cite{Pinczuk}. The present calculation
suggests that this experiment has actually not detected the
{\em lowest} branch of the collective mode, which only exists
at finite $T$ and vanishes at small $q$. On the other hand,
this lowest mode can in principle be measured through the recently
suggested experiment using evanescent field Raman scattering
\cite{KBB}.
Since the spectrum at $q \rightarrow \infty$ limit is
independent of $T$, our work also provides a justification for
the fitting of $\sigma_{yy}$ data with a constant
activation energy $\Delta$ \cite{book,Du}.

For FQH states other than the `fundamental' one, the zero $T$
calculation shows that there are more than
one roton minimum, and the number of minima corresponds to the number
of filled LLs in the composite fermion picture \cite{SH}. In Fig. 3
we show the lowest branch of the collective mode in the $\nu=3/7$ state.
At zero $T$, there are three roton minima. As $T$ is raised, the weakest
one, located at a large wavevector ($\approx 4.75\tilde{l}_{0}q$), is
smeared out first by
thermal fluctuations. The second one disappears subsequently at
higher $T$, leaving only one roton minimum at sufficiently high
temperature. This is a general feature for all the $\nu$ values
we have examined.
While the position (and the number) of roton minima has not
yet been subjected to direct experimental measurement, it does
have physical implications according to one of the proposed
scenario of FQH state--Wigner crystal transition \cite{book,GMP}.
In this picture, such a transition is directly related to the
softening of the roton minimum and the position of the (softened)
roton minimum corresponds to the reciprocal wavevector
of the Wigner crystal near the transition. Such a scenario is
not supported by the present calculation since the lattice
constant of Wigner crystal should be determined by the electron
density alone, and hence should be independent of $T$.

{\sl Summary} --
In the absence of impurity scattering, essential aspects of FQHE,
such as the quantization of Hall conductance as well as
quasi-particle charge and statistics are unaffected by
thermal fluctuations.  Within the fermion Chern-Simons
field theory approach, we show that
the magnitude of energy gap in the collective
excitation spectrum is weakly $T$ dependent, while
the number and position of roton minima are quite sensitive to $T$;
The lowest branch of the $T=0$ collective modes splits into
two when temperature is raised, where the lower one vanishes at long
wavelength limit. Our results call for more extensive
investigations on the effect of thermal fluctuations to the FQHE.

{\em Acknowledgement} --
The author thanks G.S. Canright, X. Chen, J.K. Jain and J. Quinn
for useful discussions, and G.S. Canright and X. Chen for critical
reading of the manuscript.
Valuable comments from N.E. Bonesteel, D.H. Lee, M. Ma, F.C. Zhang
and S.C. Zhang are also gratefully acknowledged.
This work was supported by the National Science Foundation
under Grant Nos. DMR-9101542 and DMR-9413057, and by the U.S. Department
of Energy
through Contract No. DE-AC05-84OR21400 administered by Martin
Marietta Energy Systems Inc..

\figure{
Figure 1. Lowest few collective modes at $\nu = 1/3$.
The width of the curve is proportional to $q^2$ times the weight
of the pole in $K_{00}$. As $T$ is raised
fron zero, each mode splits to two.
}

\figure{
Figure 2. Energy and position of the roton minimum for the
$\nu = 1/3$ state as function
of $T$. Initially, $q_{min}$ is roughly unchanged as $T$ is increased
from zero, then decreases faster as $T$ is further increased.
$\omega_{min}$ on the other hand is quite insensitive to $T$
for the whole range of temperature considered.
}

\figure{
Figure 3. Evolution of the lowest collective mode in the case of
$\nu = 3/7$. As $T$ is increased from zero, this lowest mode also
splits into two. The number of roton minima decreases as $T$ is
increased so that for sufficiently large $T$ only one minimum remains.
}

\begin{references}

\bibitem{Laughlin}
R.B. Laughlin Phys. Rev. Lett. {\bf 50} 1395 (1983).

\bibitem{book}
For a review, see, e.g., {\sl The Quantum Hall Effect},
eds. R.E.Prange and S.M.Girvin, Springer-Verlag, New York, 1990.

\bibitem{Jain}
J.K.Jain, Phys.Rev.Lett., {\bf 63}, 199 (1989); Phys.Rev.B {\bf 41},
7653 (1990); Adv. Phys. {\bf 41}, 105 (1992).

\bibitem{Du} R.R.Du , H.L.Stormer, D.C.Tsui, L.N.Pfeiffer, and K.W.West,
Phys.Rev.Lett. {\bf 70}, 2944 (1993).

\bibitem{GMP}
S.M. Girvin, A.H. MacDonald and P.M. Platzman, Phys.Rev.Lett.,
{\bf 54}, 581 (1985); Phys.Rev.B, {\bf 33}, 2481 (1986).

\bibitem{CSLG}
S.C. Zhang, H. Hansson and S. Kivelson, Phys. Rev. Lett. {\bf 62},
82 (1989); D.H. Lee and S.C. Zhang, {\sl ibid} {\bf 66}, 1220 (1991).

\bibitem{LF}
A. Lopez and E. Fradkin, Phys. Rev. B {\bf 44} 5246 (1991);
ibid, {\bf 47} 7080 (1993).

\bibitem{HLR}
B. I. Halperin, P. A. Lee, and N. Read, Phys.Rev.B {\bf 47}, 7312 (1993).

\bibitem{SH}
S.H.Simon and B.I.Halperin, Phys.Rev.{\bf B48}, 17368 (1993);
{\sl unpublished}

\bibitem{HSH}
S. He, S.H.Simon and B.I.Halperin, {\sl unpublished}

\bibitem{LZ}
L. Zhang, {\sl unpublished}.

\bibitem{Willett}
R.L. Willett et al, Phys. Rev. Lett. {\bf 65}, 112 (1990);
Phys. Rev. {\bf B47}, 7344 (1993); {\sl unpublished}.

\bibitem{RSS}
S. Randjbar-Daemi, A. Salam and J. Strathdee, Nucl. Phys. {\bf B340},
403 (1990).

\bibitem{PN}
D. Pines and P. Nozi\'{e}res, {\sl The Theory of Quantum Liquids}
{\bf I}, Benjamin (1966).

\bibitem{Fradkin}
E. Fradkin, {\sl Field Theories of Condensed Matter Systems},
Addison-Wesley (1991).

\bibitem{SK}
S.L. Sondhi and S.A. Kivelson, Phys. Rev. B {\bf 46}, 13319 (1992).

\bibitem{CQ}
X. Chen and J. Quinn, {\sl unpublished}.

\bibitem{Kohn}
W. Kohn, Phys. Rev. {\bf 123}, 1242 (1961).

\bibitem{Pinczuk}
A. Pinczuk, B.S. Dennis, L.N.Pfeiffer, and K.W.West, Phys. Rev. Lett.
{\bf 70}, 3983 (1993).

\bibitem{KBB}
A. Kuklov, A. Bulatov and J. Birman, Phys. Rev. Lett.
{\bf 72}, 3855 (1994).

\end{references}
\end{document}